\documentclass[11pt]{article}
\usepackage[latin1]{inputenc}
\usepackage[english]{babel}
\usepackage[namelimits]{amsmath}
\usepackage{amssymb}
\usepackage{amsmath}
\usepackage{amsthm}

\begin{document}
\title{Physically motivated uncertainty relations\\ at the Planck length
for an emergent\\ non commutative spacetime}
\author{{Luca Tomassini\thanks{Dipartimento di Scienze, Universit\'a di Chieti-Pescara
"G D Annunzio", Viale Pindaro 42, I-65127 Pescara, Italy and
Department of Theorethical Physics, Courant Research Centre ``Higher Order Structures'', University of G\"ottingen, G\"ottingen, Germany,  E-mail: {\tt tomassin@axp.mat.uniroma2.it}} 
$^{\dagger}$ and 
{Stefano Viaggiu\thanks{Dipartimento di Matematica,
Universit\`a di Roma ``Tor Vergata'',
Via della Ricerca Scientifica, 1, I-00133 Roma, Italy.
E-mail: {\tt viaggiu@axp.mat.uniroma2.it}.}}}}
\date{\today}\maketitle

\begin{abstract}
We derive new spacetime uncertainty relations (STUR) at the fundamental Planck length
$L_P$ from quantum mechanics and general relativity (GR), both in flat and
curved backgrounds. Contrary to claims present 
in the literature, our 
approach suggests that no minimal uncertainty appears for lengths, but instead for minimal space and four-volumes. 
Moreover, we derive a maximal absolute value for the energy density. 
Finally, some considerations on possible commutators among quantum operators implying our STUR are done.
\end{abstract}

\section{Introduction}

The idea that limitations on the localizability of spacetime events should appear at very small distances is very old and can perhaps be traced back to Wigner and Salecker \cite{t7, t8}. In a thorough investigation on the impact of quantum mechanics on the precision of measurements of distances between points and hence of the metric on curved spacetimes, the first author remarked that ``the establishment of a close network of points in space-time requires a reasonable energy density (...). However, it is not necessary to discuss this in detail because the measurement of the distances between points of the network gives more stringent requirements than the establishment of the network''.

Later, the attention shifted precisely on the conditions imposed on ``the establishment of the network'' itself, that is on the very possibility of giving an operational definition of the concept of spacetime event.
A variety of strategies were devised to attack the problem (see for example 
\cite{1,2,3,4,5,6,26q,7,8,8b,t1,t2,t3,t4,t5,t6} and the next Section for a brief discussion), but as far as we know it was Mead \cite{2} who first observed that any attempt to determine the position of an object in a spacetime 
region $\cal O$ of size comparable with the error (uncertainty) one is willing to accept entails, by the Heisenberg uncertainty principle, the localization of a non-zero amount of energy in $\cal O$ itself. It is then reasonable, Mead argued, to expect that a too high precision could trigger the formation of an event horizon representing a black hole surrounding $\cal O$, thus preventing any signal to go back to the observer and making the measurement itself meaningless. In this context, we point out that the remark of Einstein himself \cite{AA} that one cannot put an arbitrary amount of energy in a given finite 
spatial volume becomes relevant.

The relevant scale for the appearance of such effects was determined as the one of the Planck length

\begin{equation}
L_P=\sqrt{\frac{G\hbar}{c^3}}\simeq 1,6.10^{-33}{\text cm}.
\label{1}
\end{equation}

This point of view has been revived and made more precise in \cite{3} (see \cite{Piacitelli} for a nice review and introduction to more recent developments), where the following program (for flat Minkowski space) was proposed:
\begin{itemize}
\item[1)] Derive operationally meaningful uncertainty relations between coordinates of spacetime events from the requirement of gravitational stability under localization experiments.
\item[2)] Promote the above-mentioned coordinates to the status of quantum operators and commutation relations among them from which the uncertainty relations follow.
\end{itemize}
Making use of a Newtonian approximation of the gravitational potential associated to the localizing energy (see Appendix 2), the following uncertainty relations (DFR relations from now on) were obtained
\begin{eqnarray}
& &c\Delta t\left(\Delta x^1+\Delta x^2+\Delta x^3\right)\geq L^{2}_{P},\label{2}\\
& &\Delta x^1\Delta x^2+\Delta x^1\Delta x^3+\Delta x^2\Delta x^3\geq L^{2}_{P}.\label{3}
\end{eqnarray}
The authors also succeeded in making Step 2) above, thus obtaining a physically motivated model of 
a non commutative 
(see \cite{10}) quantum spacetime. Moreover, the foundations were laid down for the construction of quantum fields on it. Last but not least, it was
first observed that in all previous works the necessity of a minimal uncertainty length could be traced back to the fact that only spherically symmetric geometries are usually considered. More on this point in the next Section.

Note that equation (\ref{3}) implies that we can determine with arbitrary precision up to two spatial coordinates, say $x^1, x^2$, meaning $\Delta x^1, \Delta x^2 <\epsilon$ with $\epsilon >0$ arbitrarily small. Obviously, we have to pay the price of an arbitrarily large $\Delta x^3$. This is where we see that the extra freedom of non-spherical geometries come into action. However, if we take only $\Delta x^1<\epsilon$, then $\Delta x^2$ \textit{and} $\Delta x^3$ might well remain finite (say of order $L_P$). This would imply, again by the Heisenberg uncertainty relations and in agreement with the results of \cite{9}, that an \textit{infinite} amount of energy could be localized inside an \textit{arbitrarily small} space volume. Since event horizons have been ruled out by assumption, an unphysical naked singularity would arise.

This last observation shows the necessity of going beyond the Newtonian approximation and was the starting point of our investigations. Indeed, in this paper we take exactly the same point of view of \cite{3}, but in our case GR enters the scene through (a refinement of) the so-called \textit{hoop conjecture} (see below), first formulated by Thorne in \cite{12}. We will see in the following that in our proposed spacetime uncertainty relations the above pathologies do not occur and accordingly we will find an upper limit for the space-volume density of energy.

This paper is organized as follows. In section 2 we fix some notations and concepts.
In section 3 we review some "gedanken experimenten" present in the literature.
In section 4 we derive the STUR with a Minkowskian background metric together with some physical consequences. 
Section 5 is devoted to an attempt to generalize the STUR to a general curved background. 
Section 6 collects some final remarks and conclusions. Finally, in Appendix A we show that the approximations needed to obtain our STUR entail no change in their physical consequences, while in Appendix B we analyze those leading to relations (\ref{2}), (\ref{3}).

We conclude this introduction with two more remarks. First, we make no attempt to tackle Step 2). Second, we do not claim we provide proves, our aim is only to convince the reader by sound and hopefully robust physical arguments that our uncertainty relations are a realistic proposal: being there no chance of a direct experimental verification, they can only be tested by their {\it consequences}.

\section{Notation and preliminaries}

It is well known (see for example \cite{Landau}) that  position, momentum and energy are described in 
quantum mechanics
by self-adjoint operators acting on some Hilbert space $\cal H$, while the time $t$ is only a parameter. In what follows, the symbol $\Delta (A)$ indicates the uncertainty of the operator $A$ in some quantum state $\omega$ (for example a vector state $\omega (A)=(\phi,A\phi)$ where $\phi\in \cal H$ and $(\cdot , \cdot)$ indicates the scalar product) as defined by the formula (variance)
\begin{equation}
\Delta (A)= \Delta_{\omega} (A) = \sqrt{\omega((A -\omega (A)I)^2} = \sqrt{\omega(A^2)-\omega(A)^2}
\end{equation}
while ($t$ being a parameter) $\Delta t = t_2-t_1$ where for the sake of definiteness, $t_1$ and $t_2$ 
should be thought of as the times at which the measurements determining the region $\cal O$ are performed). The symbol $I$ of course indicates the identity operator
on $\cal H$.\footnote{For any spectral value of an operator $A$, there is a state $\omega$ with arbitrary small uncertainty: we see that an absolute lower bound in the uncertainty of one coordinate makes its description by means of an operator impossible. At the same time, in \cite{9} operators describing distances, space and four-volumes defined by \textit{independent} space-time events were defined in terms of coordinates obeying \eqref{2} and \eqref{3}. Lower bounds were found on euclidean distances and four-volumes.}
Once the average is taken, the resulting quantities behave as classical ones. Note that since the functional $\Delta$ is not linear, $\Lambda_{\mu}^{\nu}\Delta (x_{\nu})\neq \Delta (\Lambda_{\mu}^{\nu}x_{\nu})$, where $\Lambda_{\mu}^{\nu}$ is any Lorentz transformation and $x^\mu$ are the coordinates (operators) with $x^0 =t$. Thus, it is not the uncertainty relations that have to be covariant but the quantum operators they come from.

Suppose that now an observer wants to perform a measurement of position and/or time in, say, a flat Minkowsky spacetime and he is willing to accept an error $\Delta x^i$ ($i=1, 2, 3$), $\Delta t$. Then, he will have to concentrate a certain amount of matter-energy in a spacetime region $\cal O$ of corresponding size. Should this region be sufficiently small, it is reasonable to expect that quantum effects should no more be negligible. One is thus led to consider the usual Heisenberg uncertainty relations
\begin{equation}\label{7}
\Delta x^i \Delta p^j \geq \delta_{ij}\frac{\hbar}{2} \qquad\qquad\qquad i,j=1,2,3
\end{equation}
which, in our relativistic setting, should be complemented with the time-energy relations
\begin{equation}\label{6}
\Delta t \Delta E \geq \frac{\hbar}{2},
\end{equation}
and draw the conclusion that the measurement entails an energy transfer to the (dynamic) spacetime. In particular, we see that very high energies will correspond to smaller errors in time. If in addition one requires small space uncertainties, a black hole could be produced thus preventing any signal to come out from the localization region and making the measurement itself meaningless.

As remarked in the introduction, this line of reasoning seems to have a long history and was first put on quantitative grounds by Mead in 1964 \cite{2}. There, under the assumption of spherical symmetry (i.e. $\Delta R\simeq\Delta x^1 \simeq\Delta x^2\simeq\Delta x^3$), the author derived the relation
\begin{equation}\label{minlenght}
\Delta R \geq L_P
\end{equation}
and concluded from it that an absolute minimal length had to exist.

It is interesting to note that the same result was obtained in a string theoretical context (see for example \cite{Ciafaloni}) and by model-independent considerations in \cite{Maggiore}. In both cases, the starting point were the Generalized Uncertainty Relations (GUP)
\begin{equation}
\Delta R \geq\frac{\hbar}{\Delta p} + \alpha^{\prime}\Delta p
\label{me}
\end{equation}
($\alpha^{\prime}$ is a constant depending on Regge's slope and the gravitational constant $G$), from which once more under the assumption of spherical symmetry \eqref{minlenght} follows by minimization of the second member.

 Finally, we stress that the position-momentum uncertainty relations are of a completely different nature than the time-energy one. More precisely, the first ones are relations between specific averages (variances) of quantum operators while the second ones are obtained by evaluating the transition probability between states with energies differing by an amount $\Delta E$. The difficulty stems from the impossibility to associate a bona fide self-adjoint operator with time in the case of a non-trivial (i.e. lower bounded) energy spectrum, but it has been circumvented in \cite{Brunetti} by making use of the concept of operator-valued measures (see \cite{Ludwig} for a thorough discussion). The time-energy uncertainty relation (\ref{6}) was then recovered by a slight modification of the usual position-momentum argument. However, the following arguments 
should be considered independent of such details.

\section{Minimal uncertainty length in the literature: a review}

In this section we give a brief and certainly incomplete account of how the concept of minimal uncertainty is addressed in the 
literature. We already pointed out the observation first made in \cite{3}: very often the conclusion that a minimal uncertainty exists can be traced back to the (implicit or explicit) assumption that the geometry of the problem be spherical. 

We start with the papers \cite{t5,t6}. Since they were built on the ideas of \cite{t7,t8}, the main concern of the authors is limitations on the measurement of distances between different spacetimes points. However, they give a quantitative estimate of the minimal uncertainty in the localization of each single point which is obtained by invoking the Schwarzschild metric.

A wealth of localization ``gedanken experimenten'' is presented in \cite{5}, where fluctuations of the spacetime metric are analyzed making use of photons instead of target massive particles (together with some considerations on the Einstein equations). We will do the same in the following. In the first experiment a microscope-type construction is presented, namely a photon scattered by a target particle. Here, the Newtonian gravitational interaction between 
the photon and the target particle is calculated together with the suggested modification of the Heisenberg
relation leading to relation (\ref{me}) without invoking black hole physics. In any case, the Newtonian
gravitational field is given in the form $\sim\frac{1}{r^2}$ and so spherical symmetry is
implicitly assumed. The same assumption affects the description of the time resolution of synchronizing clocks in terms of the 
gravitational field generated by photons: a time-time component of a spherically symmetric metric is introduced. 
In the second example, a minimal uncertainty length is analyzed within a path integral formulation of
quantum gravity. At a certain point the author denotes with $\ell$ "the radius of the region" and
so, once again, spherical symmetry comes on stage. In \cite{t1} another interesting localizing experiment based on clocks and on the Heisenberg uncertainty principle is presented and an uncertainty relation involving the volume $\Delta V$ of the region to be localized and the time coordinate is found. The author denotes $\Delta V={(\Delta x)}^3$ and imposes that the "Schwarzschild radius
should not exceed the uncertainty $\Delta x$": thus, the condition 
$\Delta x=\Delta y=\Delta z$ is implicitly assumed. Even the minimal uncertainty found in \cite{t2} using the GUP (\ref{me}) is obtained using a spherical black hole argument. More references presenting arguments plagued by the assumption of spherical geometry can be found in \cite{Piacitelli}.

The necessity of considering more general geometries is however clearly stressed in \cite{8b}, where an interesting localizing 
experiment using an interferometer is proposed. The authors provide sound arguments to show that the uncertainty in one single coordinate can be reduced to arbitrarily small values, but are only able to provide quantitative results in a spherical case where a minimal uncertainty is once again found.

We conclude this section by mentioning that not only uncertainties in coordinates and distances have been considered in the literature. For example, much in the spirit of the investigations of Wigner himself, in \cite{t4}, \cite{t3} the main interest are the quantum fluctuations of the metric. Spherical symmetry is assumed in \cite{t3}, while in \cite{t4} only asymptotic flatness of static spacetime is needed. In both papers a Heisenberg-type uncertainty relation is found between the metric and a ``conjugate'' quantity, so that there is no minimal uncertainty.

\section{Minkowskian background}

\subsection{Spherical case}
As it should now be clear, localization in some spacetime region $\cal O$ will mean producing a (quantum) state confined there. We start by considering an experiment performed by an observer in a flat Minkowskian spacetime. It should be no surprise that in what we call spherical case (that is $\Delta x^1 \simeq \Delta x^2 \simeq \Delta x^3 \simeq \Delta R /\sqrt{3}$, $\Delta R$ being the radius of the localization sphere) a completely satisfactory answer can be given. We start by presenting a separate treatment of this situation.

The metric outside the space volume $V^{(3)}$ where an energy $\Delta E$ is concentrated is
\begin{equation}
ds^{2}_{exp}=\frac{dr^2}{\left(1-\frac{2G\Delta E}{c^4 r}\right)}+r^2d{\Omega}^2-
\left({1-\frac{2G\Delta E}{c^4 r}}\right)dt^2.\label{4}
\end{equation}
Of course, we set $\Delta E=c^2 \Delta M$, $\Delta M$ being then the ADM mass-energy of the experiment. 
Since the configuration is spherical, gravitational radiation is absent so that the exterior metric is static. Note that, due to spherical symmetry, this will be true irrespective of the duration $\Delta t$ of the experiment, where of course $t$ is the time measured by an observer at spatial infinity.
Making use of the Komar integral, it is clear that for any $r>\Delta R$ the mass-energy of the experiment is simply $\Delta M$. From (\ref{4}), the condition that no black hole is formed reads
\begin{equation}
\frac{G\Delta E}{c^4}<\frac{\Delta R}{2}.
\label{5}
\end{equation}
As remarked above, our observer measures the time $t$ and the energy in terms of it. He can therefore use the Heisenberg relation (\ref{6}) to estimate the {\it average} energy $\Delta E$ (recall the discussion in section 2)
of the configuration, so that condition (\ref{5}) becomes
\begin{equation}
c\Delta R\Delta t\geq L^{2}_P.
\label{8}
\end{equation}
This is the time-space uncertainty relation.

Since $\Delta E \geq c\Delta p$, from (\ref{7}) and condition (\ref{5}) we obtain
\begin{equation}
\Delta R\Delta x^1\Delta x^2\Delta x^3\geq L^{2}_{p}
\sqrt{{(\Delta x^1)}^2{(\Delta x^2)}^2+{(\Delta x^2)}^2{(\Delta x^3)}^2+
{(\Delta x^1)}^2{(\Delta x^3)}^2}\label{9}.
\end{equation}
In the spherical case, this reads $\Delta x^i\geq L_P, i=1,2,3$, i.e. a minimal 
uncertainty length appears
in agreement with (\ref{minlenght}) and (\ref{me}).
Furthermore, in this particular situation, our results agree with the ones in (\ref{3}).

\subsection{General case}

The main problem when considering more general geometries than the spherical one is the absence of general theorems guaranteeing that no black holes are formed, independently of the dynamics governing the energy-matter motion.
In fact solving the global Cauchy problem remains a formidable task at the present day, apart from
the remarkable work of Christodoulou \cite{C1,C2} about spherical collapse in 
the presence of classical scalar fields.\\
For this reason, we resort to the so called hoop conjecture (see \cite{12,13,14,15,16,17,18,19}). In its original version, it states that

\begin{itemize}
\item horizons form when and only when a mass $M$ is concentrated into a region whose circumference
in every direction satisfies $C\leq 4\pi M\frac{G}{c^2}$.
\end{itemize}
For example, this would imply that a mass confined in a cylinder with a small base radius but a sufficiently large height would 
not produce an horizon. The statement is purposely imprecise. As the mass $M$ we could take the ADM mass or
a quasi-local one such as the Tolmann mass \cite{20}. The study of event horizon formation is very
complicated since it is a global concept available only to "omniscient" observers (see \cite{14}). For these reasons,
in the literature, geometrical and more tractable
objects as apparent horizons and trapped surfaces (see \cite{w1,w2,w3}) are often used.
In fact, the formation of trapped surfaces and apparent horizons represents an
indication of the formation of black holes, via the cosmic censorship conjecture \cite{w4}.\\
In any case the hoop conjecture can also be stated in terms of the surface of the horizon $A_H$ and the mass $M$, according to the so called Penrose's isoperimetric-type conjecture (see \cite{15,21}). For a stationary black hole we have \cite{13}
\begin{equation}
A_E<A_H \leq 16\pi \frac{G^2}{c^4}M^2,
\label{10}
\end{equation}
$A_E$ being the apparent horizon. As a result, the existence of an apparent horizon implies the existence of
an event horizon enclosing it. More generally, denoting
with $A_i$  the minimal area of a surface enclosing an apparent horizon, 
Penrose's isoperimetric conjecture states that
\begin{equation}
A_i\leq 16\pi \frac{G^2}{c^4}M^2.
\label{100bb}
\end{equation}
To ensure that no black hole is formed,
we need to invert inequality (\ref{100bb}). 
Furthermore, we study situations in which the mass-energy $M$ is localized within a region of area $A$
with $A\geq A_i$.
As a result, 
we can with no loss of generality consider the inequality
\begin{equation}
A \geq 16\pi \frac{G^2}{c^4}M^2,
\label{11}
\end{equation}
so that {\it in no case} the formation of an event horizon enclosing the localizing 
area $A$ is possible. 
This is our starting point. 
By considering a surface $\Delta A$ enclosing a mass $\Delta M$ and again putting $\Delta E=c^2 \Delta M$, we have
\begin{equation}
\Delta A \geq 16\pi\frac{G^2}{c^8}{(\Delta E)}^2.
\label{12}
\end{equation}
Suppose now the localization region is a box with edges $\Delta x^1,\Delta x^2,
\Delta x^3$. Then
\begin{equation}
\Delta A \sim \frac{\Delta x^1\Delta x^2+\Delta x^1\Delta x^3+\Delta x^2\Delta x^3}{{\alpha}^2},
\label{13}
\end{equation}
where the normalization constant $\alpha$ will be determined later.\\
Concerning the energy $\Delta E$, from equation (\ref{6}) we again have
\begin{equation}
\Delta E\geq\frac{\hbar}{2\Delta t}.
\label{14}
\end{equation}
Furthermore, since the localization experiment is performed by concentration of mass-energy, we have
\begin{equation}
{(\Delta E)}^2\geq c^2
\left[{(\Delta p_{x^1})}^2+{(\Delta p_{x^2})}^2+{(\Delta p_{x^3})}^2\right],
\label{15}
\end{equation}
and from (\ref{7}) we deduce that
\begin{equation}
{(\Delta E)}^2\geq\frac{c^2{\hbar}^2}{4}\left[{\left(\frac{1}{\Delta x^1}\right)}^2+
{\left(\frac{1}{\Delta x^2}\right)}^2+{\left(\frac{1}{\Delta x^3}\right)}^2\right].
\label{16}
\end{equation}
Note that inequality (\ref{16}) holds for any type of localizing mass-energy.  
From (\ref{13}), (\ref{14}) and (\ref{12}) we obtain
\begin{equation}
{(c\Delta t)}^2\left(\Delta x^1\Delta x^2+\Delta x^1\Delta x^3+\Delta x^2\Delta x^3\right) \geq
4\pi{\alpha}^2L^{4}_{P}.
\label{17}
\end{equation}
Comparison with the spherical case gives $4\pi {\alpha}^2=1$. We can use inequality (\ref{16}) to obtain the space-space uncertainty relation. To this purpose first notice that the following inequality holds
\begin{equation}
{\left(\frac{1}{\Delta x^1}+\frac{1}{\Delta x^2}+\frac{1}{\Delta x^3}\right)}^2\geq
{\left(\frac{1}{\Delta x^1}\right)}^2+
{\left(\frac{1}{\Delta x^2}\right)}^2+{\left(\frac{1}{\Delta x^3}\right)}^2. \label{18}
\end{equation}
Estimating the energy $\Delta E$ by the left hand side of (\ref{18}) we obtain a less precise result, i.e. we ask that the area $\Delta A$ be greater 
than a quantity which in turn is greater than the energy. Nevertheless, 
we argue in Appendix A that this causes no change in the physical content of the uncertainty relation. Summing up, our final expression for the STUR is
\begin{eqnarray}
{(c\Delta t)}^2\left(\Delta x^1\Delta x^2+\Delta x^1\Delta x^3+\Delta x^2\Delta x^3\right) \geq
L^{4}_{P},\qquad\quad\label{19}\\
{(\Delta x^1)}^2 {(\Delta x^2 )}^2 {(\Delta x^3 )}^2 \geq L^{4}_{P}
\left(\Delta x^1\Delta x^2+\Delta x^1\Delta x^3+\Delta x^2\Delta x^3\right).\label{20}
\end{eqnarray}
Furthermore, it easy to see that making use of the following crude estimates for 
(\ref{19}) and (\ref{20}) respectively
\begin{eqnarray}
(a+b+c)^2\geq ab+bc+ac, \label{21}\\
(ab+bc+ac)^3\geq a^2 b^2 c^2,\quad\label{22}
\end{eqnarray}
with $a=\Delta x^1, b=\Delta x^2, c=\Delta x^3$,
we get the DFR relations (\ref{2}) and (\ref{3}).

\subsection{Some physical consequences}

To start with, rewrite (\ref{20}) in the form
\begin{equation}
\Delta x^1\Delta x^2\Delta x^3 \geq L^{4}_{P}\left(\frac{1}{\Delta x^1}+\frac{1}{\Delta x^3}+\frac{1}{\Delta x^3}\right).
\end{equation}
If $\Delta x^1<\epsilon$ ($\epsilon>0$), the other uncertainties being arbitrary, we obtain
\begin{equation}
\Delta x^1\Delta x^2\Delta x^3\geq \frac{L^{4}_{P}}{\epsilon}.
\label{23}
\end{equation}
We see then that if an uncertainty on a single spatial coordinate is arbitrarily small, the spatial volume must increase accordingly. Not only no minimal uncertainty length arises, but localization of an infinite energy in a finite volume is forbidden.\\
The same conclusion can be drown in the case $\Delta x^1<\epsilon_1, \Delta x^2<\epsilon_2$ ($\epsilon_1,\epsilon_2 >0$), but now we also see that there is no minimal surface in the sense that the product of two linear uncertainties can be made arbitrarily small. In fact, we have
\begin{equation}
\Delta x^1\Delta x^2\Delta x^3\geq L^{4}_{P}\left(\frac{1}{\epsilon_1}+\frac{1}{\epsilon_2}\right).
\label{24}
\end{equation}
Moreover, by taking the third power of (\ref{20}) and estimating the resulting right-hand side by (\ref{22}) we easily 
obtain
\begin{equation}
\Delta x^1\Delta x^2\Delta x^3\geq L^{3}_{P},\label{26}
\end{equation}
i.e. a lower bound for the space volume appears. Finally, (\ref{19}) and (\ref{20}) together imply
\begin{equation}
c\Delta t\Delta x^1\Delta x^2\Delta x^3\geq L^{4}_{P}.
\label{25}
\end{equation}
meaning that an analogous result holds for the four-volume. On the other hand, if we put $\rho\Delta x^1\Delta x^2\Delta x^3=\Delta M$ ($\rho$ being thus the density) in (\ref{11}) and take advantage of (\ref{12}) and (\ref{20}) we obtain
\begin{equation}
\rho<\frac{1}{4\sqrt{\pi}}\frac{c^2}{GL_P^2}\sim {\rho}_{P},
\label{27}
\end{equation}
where ${\rho}_{P}$ indicates Planck's density. Thus, independently of the geometry of the localization zone, an upper limit for the density appears. In view of (\ref{26}) this should be no surprise, but we stress that GR alone cannot account for this result: to rule out singular distributions of matter-energy one needs quantum mechanics.
Concerning relation (\ref{19}), when the "spatial" part 
$\left(\Delta x^1\Delta x^2+\Delta x^1\Delta x^3+\Delta x^2\Delta x^3\right)$
goes to infinity, $\Delta t$ can be made arbitrarily small.
Conversely, when $\Delta x^1\sim\Delta x^2\sim\Delta x^3\sim L_p$, we have $c\Delta t\geq L_p$, showing that,
limited to the spherical case, we have a minimum uncertainty also in the time coordinate.\\
Our next considerations concern the possible underlying quantum structure. The determination of commutators among quantum operators that could give rise to our STUR is certainly not a simple task. Nevertheless, some partial conclusions can be drawn thanks to the fact that in our approach the DFR relations appear as an approximation to the STUR themselves and moreover provide exactly the same result in the spherical case. We look at this fact as a hint that the commutation relations obtained in \cite{3} could be viewed as a limiting case of the ones giving rise to the STUR.\\ 
In view of the generalization to curved backgrounds, we conclude this section by the following remarks. Introducing the surfaces
\begin{equation}
\Delta A^{0}={(c\Delta t)}^2,\; \Delta A^{1}=\Delta x^2\Delta x^3,\;\Delta A^{2}=\Delta x^1\Delta x^3,\; \Delta A^{3}=\Delta x^1\Delta x^2.
\label{28}
\end{equation}
the STUR (\ref{19}) and (\ref{20}) become
\begin{eqnarray}
\Delta A^{0}\left(\Delta A^{3}+\Delta A^{1}+\Delta A^{2}\right)\geq L^{4}_{P},\qquad\quad\label{29}\\
\Delta A^{3}\Delta A^{1}\Delta A^{2}\geq L^{4}_{P}\left(\Delta A^{3}+\Delta A^{1}+\Delta A^{2}\right).\label{30}
\end{eqnarray}
Moreover, by making use once more of (\ref{22}), we obtain
\begin{equation}
\Delta A^{3}\Delta A^{2} +\Delta A^{3}\Delta A^{1}+\Delta A^2 \Delta A^{1}\geq L^{4}_{P},
\label{30b}
\end{equation}
Note that, together with (\ref{29}), this is exactly the same structure of the DFR relations, albeit in terms of the surfaces $\Delta A^{0}, \Delta A^{1}, \Delta A^{2}, \Delta A^{3}$. We believe that this might be more than a simple formal manipulation, since we have seen that GR (more precisely condition (\ref{12})) and quantum mechanics seem to lead naturally to terms proportional to $L^{4}_{P}$. More importantly, the commutators 
$[A^{\alpha},A^{\beta}]$ (where $\alpha, \beta=0,\dots,3$)
between the quantum
operators $A^{\alpha}$ are known, being formally identical 
to the commutators between the spacetime coordinates present in \cite{3}. The next step would be to find explicit expressions for the coordinate operators $t,x,y,z$ in terms of the $A^{\alpha}$'s and calculating the corresponding commutators.

\section{Curved background: an attempt}

The extension of our results to the case of a general curved background presents one main difficulty. Of course, we still have condition (\ref{12}) at our disposal, but unfortunately we cannot use Heisenberg relations (\ref{6}) 
and (\ref{7}) any more. The main problem is that, thanks to the equivalence principle, there
is not a simple manner to isolate the energy contribution inside 
the localizing "box" (the experiment) from the
rest of the mass-energy distribution, i.e for a non-isolated system. However, in the literature 
(see \cite{26,27,28,29} and for a comprehensive review \cite{30} and references therein) the 
concept of quasi-local mass has been introduced
to measure the energy of a system within a closed compact
orientable space-like two-surface by means
of a energy-momentum four-vector depending on the first fundamental form on the two-surface itself,
the second fundamental form and null normals which define a preferred basis for the fibers of the 
normal bundle. It is a common opinion that the most promising was proposed in \cite{29} by means of
the Hamiltonian formulation of GR.
Obviously, the most important  requirements for a quasi-local mass are as follows.
\begin{enumerate}
\item The ADM and Bondi mass should be recovered respectively at spatial and null infinity.
\item When the surface degenerates to a point, correct limits must be obtained.
\item The quasi-local mass must be non-negative and zero for any space-like two surface in a
flat Minkowskian background.
\end{enumerate}
The conditions above are not trivial to fulfill. We do not enter in more details and
thereafter denote with the symbol $E_{ql}$ the quasi-local energy.\\
Furthermore, for our purposes, it seems to be reasonable to use the formalism introduced in 
\cite{23} and \cite{24}.
Then, $g_{\mu\nu}$ being the background metric, we define
\begin{eqnarray}
h_{\mu\nu}=g_{\mu\nu}+{\gamma}_{\mu\nu},\;
{\gamma}^{0}=\frac{1}{\sqrt{|g_{00}|}},\;{\gamma}^{i}=0,\;
{\gamma}_i=\frac{g_{0i}}{\sqrt{|g_{00}|}},\label{42}\\
ds^2=d{\sigma}^2-c^2dT^2,\qquad\qquad\qquad\label{43}
\end{eqnarray}
where
\begin{equation}
d{\sigma}^2=h_{ik}dx^i dx^k,
\label{44}
\end{equation}
with $T$ given by 
\begin{equation}
T=\int\sqrt{|g_{00}|}dt+\int\frac{g_{0i}}{c\sqrt{|g_{00}|}}dx^i,\;\;\;i=1,2,3.
\label{39}
\end{equation}
In this way the line element is written only in terms of observable
quantities, i.e. $T$ and $\sigma$. As a consequence, we propose to substitute inequality
(\ref{6}) for a flat background with
\begin{equation}
{\Delta}E_{ql}\Delta T\geq\frac{\hbar}{2}.
\label{z1}
\end{equation}
Expression (\ref{z1}) is generally well defined and reduces to inequality (\ref{6}) in a flat
background.
As a consequence, using (\ref{z1}) and (\ref{12}) we can write the time-space uncertainty relation as
\begin{equation}
\Delta A^{00}\Delta A \geq L^{4}_{P},\;\;\;\Delta A^{00}={(c\Delta T)}^2,
\label{40}
\end{equation}
where $\Delta A$ is now a surface depending on the
non-flat background metric (\ref{43}) and cannot be expressed in the simple form (\ref{13}).\\
The space-space relation is much more involved and we have not been able to find a reasonable generalization 
for a curved background. However, some interesting inequalities can 
be suggested.\\
First of all,
if we consider a stationary background, $h_{ik}$ does not depend
on the time $T$ and as a result, we can write equations (\ref{19}) and
(\ref{20}) in a form more appropriate for a curved background without any explicit
reference to the spatial coordinates. We obtain
\begin{eqnarray}
& &c\Delta T\Delta V^{(3)} \geq L^{4}_{P},\label{45}\\
& &{(\Delta V^{(3)})}^2\geq L^{4}_p\;\Delta A. \label{450}
\end{eqnarray}
Obviously, expressions (\ref{45}) and (\ref{450})
reduce to (\ref{19}) and (\ref{20}) in a Minkowskian background by writing the 
spatial volume and the total surface of the localizing "box" in terms of Cartesian coordinates.
We regard expressions (\ref{40}), (\ref{45}) 
and (\ref{450}) as an indication that proper times, surfaces and volumes may be the natural object to be quantized.\\
Taking into account that inequality (\ref{26}) must still hold in a 
\textit{stationary} general background ($\Delta V^{(3)}\geq L^{3}_P$),
 equation (\ref{45}) together with (\ref{40}) entails an upper limit to the proper density $\rho$. In fact, in analogy with the flat case, we can write the proper mass (see \cite{18}) $\Delta M_p$ (the mass inside the proper three volume 
$\Delta V^{(3)}$) as $\rho\Delta V^{(3)}=\Delta M_p$. As a consequence, recalling (\ref{12}) we see that inequality (\ref{27}) is still valid independently of the geometry of the configuration. Once more, 
we stress that this result is a consequence of \textit{both} GR
and quantum mechanics.\\ 
In any case, for a general non-stationary background, it seems reasonable to
take advantage of the results in (\ref{25}) to obtain 
\begin{equation}
\Delta V^{(4)} \geq L^{4}_{P},
\label{41}
\end{equation}
$\Delta V^{(4)}$ being the invariant four-volume. 
Our method allows no further generalizations, since in a non-stationary background the four-volume (\ref{41})
cannot be put in the form (\ref{45}), because of the time dependence of the metric $h_{ik}$ .\\
Conversely, if we adopt inequality (\ref{45}) also for a non-stationary background, expression
(\ref{41}) cannot be directly inferred.\\
Summarizing, for a stationary background spacetime we use (\ref{40}), (\ref{45}) 
and (\ref{450}). Obviously they are not independent. In fact, starting with (\ref{450}) and using
(\ref{40}), we obtain inequality (\ref{45}). For a general time-dependent spacetime, we use 
(\ref{40}) and (\ref{41}).\\ 
Let us conclude this section with a remark.
Recently, in \cite{222} it has been advocated that energy can be localized in GR by means of a  
generalization of the Tolman mass (\cite{20}) to a generic non-static background. The expressions  for the energy and momentum proposed there are
\begin{eqnarray}
& &\tilde{E}=\frac{c^4}{4\pi G}\int R^{\mu}_{\nu}u^{\nu}u_{\mu}\sqrt{-g}d^4 x,\;\;\;
\mu,\nu=(0,1,2,3),\label{35}\\
& &{\tilde{p}}^{\alpha}=\frac{c^4}{4\pi G}\int R^{\alpha}_{\nu}u^{\nu}\sqrt{-g}d^4 x,\;\;\;
\alpha=(1,2,3),\label{36}
\end{eqnarray}
where $u^{\nu}$ stands for the four-velocity of the observer. The main idea is that the energy is encoded by the 
Ricci tensor $R_{\mu\nu}$. We stress that if (\ref{35}) and (\ref{36}) are correct, gravitational waves 
do not carry energy.\\
For a free-falling observer we have $u^{\nu}=u_{\nu}=(1,0,0,0)$. Since (\ref{35})
is an invariant four-scalar, it is clear that he/she computes the 
Tolman mass (multiplied by its proper time), i.e. the mass ``without gravity''. As a consequence, in \cite{222} the following uncertainty relations are proposed
\begin{eqnarray}
\Delta E\Delta t\geq\frac{\hbar}{2}\;\;\rightarrow\;\;\Delta\tilde{E}\geq\frac{\hbar}{2},\;\;\label{37}\\
\Delta p^{\alpha}\Delta x^{\alpha}\geq \frac{\hbar}{2}\;\;\rightarrow\;\;
\Delta{\tilde{p}}^{\alpha}\geq\frac{\hbar}{2}.\label{38}
\end{eqnarray}
Thanks to the observations above concerning a free-falling observer, to impose condition (\ref{12}) to (\ref{37}) we can divide the quantity $\Delta\tilde{E}$ by the chrono-invariant time given by (\ref{39}).
As a consequence, using (\ref{37}) the time-space uncertainty relation becomes the one given by
(\ref{40}). For the space-space relation,
we could use expression
(\ref{38}), but this cannot be written in the form (\ref{16}) since the norm of a vector is given in terms
of the background metric. Hence, we are forced to use again the arguments leading to (\ref{45})
and (\ref{41}).

\section{Conclusions}

In this paper we obtained the STUR from Quantum Mechanics and classical GR. More precisely, we made use of the Heisenberg uncertainty relations and of the observation that the formation of black holes due to concentration of energy makes the measurement of position operationally meaningless. As advocated in \cite{3}, the STUR then follow from a sound physical input.

To start with, we obtained STUR in a flat background. Then, we have generalized them to general curved backgrounds
with particular attention to the stationary case.
Our STUR appear completely physically reasonable, in the sense that an infinite energy can never be localized in a finite volume.
Furthermore, in agreement with \cite{3}, we found no minimal uncertainty while a natural lower bound for the Lorentz invariant (or general invariant in the case of GR) four-volume appears together with an upper limit for the space-volume density. We stress that our results coincide, in the case of a spherical geometry, with the ones obtained
by making use of the GUP, that is, by allowing the Heisenberg relations to be modified by gravity.\\
Some indications were given for the possible underlying quantum structure, 
but we were not able to find the commutators implying our STUR. This will be the subject of a future work.

\section*{Appendix A} 
The exact form of our STUR (i.e. without making use of the approximation (\ref{18})) is
\begin{eqnarray}
{(c\Delta t)}^2\left(\Delta x^1\Delta x^2+\Delta x^1\Delta x^3+\Delta x^2\Delta x^3\right)\geq
L^{4}_{P},\label{B1}\qquad\\
{(\Delta x^1)}^2{(\Delta x^2)}^2{(\Delta x^3)}^2
\left(\Delta x^1\Delta x^2+\Delta x^1\Delta x^3+\Delta x^2\Delta x^3\right)\geq \nonumber\\
\geq L^{4}_{P}\left({(\Delta x^1)}^2{(\Delta x^2)}^2+{(\Delta x^2)}^2{(\Delta x^3)}^2+
{(\Delta x^1)}^2{(\Delta x^3)}^2\right).\label{B2}
\end{eqnarray}
First of all observe that we can write (\ref{B2}) as
\begin{eqnarray}
\Delta x^1\Delta x^2\Delta x^3 \geq L^{4}_{P}(\Delta x^1\Delta x^2 +\Delta x^2\Delta x^3 +\Delta x^3\Delta x^1)^{-1}\times \nonumber\\
\times\left(\frac{\Delta x^1\Delta x^2}{\Delta x^3}+\frac{\Delta x^2\Delta x^3}{\Delta x^1}+\frac{\Delta x^3\Delta x^1}{\Delta x^2}\right). \qquad\qquad
\label{B3}
\end{eqnarray}
Consider now the case $\Delta x^1<\epsilon$ ($\epsilon>0$), $\Delta x^2,\Delta x^3>\Delta x^1$. Then
\begin{equation}
\Delta x^1\Delta x^2\Delta x^3 \geq \frac{L^{4}_{P}}{3\Delta x^2\Delta x^3}\times\frac{\Delta x^2\Delta x^3}{\Delta x^1}=\frac{L^{4}_{P}}{3\Delta x^1},
\end{equation}
in agreement with (\ref{23}). For $\Delta x^1<\epsilon_1, \Delta x^2<\epsilon_2$ ($\epsilon_1>0, \epsilon_1>0$) and $\Delta x^3>\Delta x^2, \Delta x^2$ we have, this time in agreement with (\ref{24}),
\begin{eqnarray}
\quad \Delta x^1\Delta x^2\Delta x^3\geq \frac{L^{4}_{P}}{2(\Delta x^2\Delta x^3+\Delta x^3\Delta x^1)}\times\left(\frac{\Delta x^2\Delta x^3}{\Delta x^1}+\frac{\Delta x^3\Delta x^1}{\Delta x^2}\right)= \nonumber \\
=\frac{L^{4}_{P}}{2}\left(\frac{\Delta x^2}{\Delta x^1(\Delta x^1+\Delta x^2)}+\frac{\Delta x^1}{\Delta x^2(\Delta x^1+\Delta x^2)}\right)\geq \qquad\qquad\nonumber \\
\geq\frac{L^{4}_{P}}{2}\left(\frac{\Delta x^2}{(\Delta x^1+\Delta x^2)^2}+\frac{\Delta x^1}{(\Delta x^1+\Delta x^2)^2}\right)= \qquad\qquad\quad\nonumber \\
=\frac{L^{4}_{P}}{2}\left(\frac{1}{(\Delta x^1+\Delta x^2)}\right)\geq   \frac{L^{4}_{P}}{2(\epsilon_1+\epsilon_2)}.\qquad\qquad\qquad\qquad
\end{eqnarray}
The case $\Delta x^i<\epsilon_i$ and $\epsilon_i$ arbitrarily small ($i=1,2,3$) is forbidden by (\ref{B2}). To estimate the three-volume first note that we can suppose with no loss of generality that $\Delta x^1\Delta x^2=\max \{\Delta x^1\Delta x^2, \Delta x^2\Delta x^3, \Delta x^1\Delta x^3 \}$. Then from (\ref{B2}) we obtain
\begin{eqnarray}
{(\Delta x^1)}^2{(\Delta x^2)}^2{(\Delta x^3)}^2\geq \frac{L^{4}_{P}}{3}\Delta x^1 \Delta x^2 \geq \nonumber \\
\geq \frac{L^{4}_{P}}{9}\left(\Delta x^1\Delta x^2+\Delta x^1\Delta x^3+\Delta x^2\Delta x^3\right).
\end{eqnarray}
By taking the third power of both sides and using (\ref{22}), we arrive at
\begin{equation}
\left[{(\Delta x^1)}^2{(\Delta x^2)}^2{(\Delta x^3)}^2\right]^3\geq \frac{L^{12}_{P}}{9^3}{(\Delta x^1)}^2{(\Delta x^2)}^2{(\Delta x^3)}^2,
\end{equation}
which gives
\begin{equation}
\Delta x^1 \Delta x^2 \Delta x^3\geq \frac{L^{3}_{P}}{3\sqrt{3}}.
\end{equation}
Finally, for the four-volume we obtain
\begin{eqnarray}
{(c\Delta t)}^2{(\Delta x^1)}^2{(\Delta x^2)}^2{(\Delta x^3)}^2\geq \qquad\qquad\quad \label{B4}\\
L^{8}_{P}
\frac{\left({(\Delta x^1)}^2{(\Delta x^2)}^2+{(\Delta x^2)}^2{(\Delta x^3)}^2+
{(\Delta x^1)}^2{(\Delta x^3)}^2\right)} 
{{\left(\Delta x^1\Delta x^2+\Delta x^1\Delta x^3+\Delta x^2\Delta x^3\right)}^2}.\nonumber
\end{eqnarray}
and by the same assumption then for the three-volume
\begin{eqnarray}
\frac{\left({(\Delta x^1)}^2{(\Delta x^2)}^2+{(\Delta x^2)}^2{(\Delta x^3)}^2+
{(\Delta x^1)}^2{(\Delta x^3)}^2\right)}
{{\left(\Delta x^1\Delta x^2+\Delta x^1\Delta x^3+\Delta x^2\Delta x^3\right)}^2}\geq \nonumber \\
\geq \left(\frac{\Delta x^1\Delta x^2}{\Delta x^1\Delta x^2+\Delta x^1\Delta x^3+\Delta x^2\Delta x^3}\right)^2 \geq \frac{1}{9}, \qquad\nonumber
\end{eqnarray}
so that
\begin{equation}
c\Delta t \Delta x^1 \Delta x^2 \Delta x^3 \geq \frac{L^{4}_{P}}{3}.
\end{equation}

\section*{Appendix B}
In \cite{3} the condition that no trapped surface is formed is $-\phi<1$, $\phi$ being of course the gravitational potential in the Newtonian approximation. By considering a constant energy density $\rho(-t)$ at negative times $x^0=-t$ over a volume centered at the origin with sides $\Delta x^1+t, \Delta x^2+t, \Delta x^3+t$, the authors obtain $\phi\sim\Delta E\times I$, with
\begin{equation}
I=-\int_{0}^{\infty}\frac{rdr}{(r+\Delta x^1)
(r+\Delta x^2)(r+\Delta x^3)}.
\label{A1}
\end{equation}
Notice that
\begin{equation}
-I=\frac{\Delta x^1\Delta x^2\ln\frac{\Delta x^1}{\Delta x^2}+
\Delta x^1\Delta x^3\ln\frac{\Delta x^3}{\Delta x^1}+
\Delta x^2\Delta x^3\ln\frac{\Delta x^2}{\Delta x^3}}{(\Delta x^1-\Delta x^3)(\Delta x^2-\Delta x^3)(\Delta x^1-\Delta x^2)}.
\label{A3}
\end{equation}
Putting $\Delta A_{DFR}=I^{-2}$, condition (\ref{A1}) becomes
\begin{equation}
{(\Delta E)}^2<\frac{c^8}{G^2}\Delta A_{DFR}.
\label{A2}
\end{equation}
Expression (\ref{A2}) is now comparable (apart from an inessential constant) with our starting condition 
(\ref{12}).
Much in the spirit of \cite{3}, we will take $\Delta E$ as given by (\ref{14}) and (\ref{16}).

To begin with, it is easy to see that in the spherical case $\Delta A_{DFR}\sim {(\Delta x^1)}^2$ so that (\ref{A3}) and (\ref{13}) differ only by a positive constant thus giving rise to the same physical consequences.
For $\Delta x^1=\Delta x^2>>\Delta x^3$ (a large disk), the result is $\Delta A_{DFR}\sim 
{(\Delta x^1)}^2$ and $\Delta A\sim {(\Delta x^1)}^2$ (where of course $\Delta A$ is given in (\ref{13})), so that in
this case too the two methods agree. However, for a cylindrical geometry with $\Delta x^1>>\Delta x^2\sim\Delta x^3$ we find 
$\Delta A_{DFR}\sim\Delta A\sim {(\Delta x^1)}^2$ and
we see that $\Delta A_{DFR}$ is an overestimation compared to our calculations. In a sense, the approximation considered in \cite{3} amounts to identifying a cylinder with the sphere containing it. Of course, a greater $\Delta A_{DFR}$ means a smaller $|\phi|$ so that the preceding overestimation appears to be a consequence of the weak field approximation.

We conclude with the following remark. The fact that the use of the weak field approximation can give sound results should be no surprise. For example, if we take the spherical static metric
\begin{equation}
ds^2=-e^{2\phi(r)}c^2 dt^2+e^{-2\psi(r)}dr^2+r^2d{\Omega}^2,
\label{A4}
\end{equation} 
and recall that in the weak approximation $\phi(r)$ becomes the Newtonian potential, we obtain
\begin{equation}
\phi=\ln\sqrt{1-\frac{2GM}{c^2}}\simeq -\frac{GM}{c^2 r}.
\end{equation}
Since (remember that in the weak field approximation $|\phi|<<1$) the condition for no trapped surfaces
becomes $\frac{GM}{c^2 r}<1$, it is wrong only by a factor 2.

\section*{Acknowledgements} 
We would like to thank Sergio Doplicher for suggesting the problem and his constant
encouragement, Gerardo Morsella and Nicola Pinamonti for several interesting discussions and suggestions.
LT  would also like to thank
Sebastiano Carpi for many useful hints. 
Research by LT has supported by GNAMPA-INDAM and the German Research Foundation (Deutsche Forschungsgemeinschaft (DFG)) through the Institutional Strategy of the University of G\"ottingen.
Finally, we would like to thank the referees for their observations and suggestions.

\end{document}